\documentclass[10pt,a4paper]{article}
\usepackage{dsfont,amsmath,amssymb,graphicx,cite}

\begin{document}

\noindent{\Large\textbf{Fragment-synthesis-based multiparty cryptographic key distribution over a public network}}

\vspace{5pt}
\noindent\textbf{\sc{Wen-Kai Yu$^{1,2,*}$, Ya-Xin Li$^{1,2}$, Jian Leng$^{1,2}$, and Shuo-Fei Wang$^{1,2}$}}

{\raggedright\footnotesize\textit{$^1$Center for Quantum Technology Research, School of Physics, Beijing Institute of Technology, Beijing 100081, China}}\\
{\raggedright\footnotesize\textit{$^2$Key Laboratory of Advanced Optoelectronic Quantum Architecture and Measurement of Ministry of Education, School of Physics, Beijing Institute of Technology, Beijing 100081, China}}

\renewcommand{\thefootnote}{\fnsymbol{footnote}}
\footnotetext[1]{\footnotesize{Corresponding author: W.-K. Yu

\ \ E-mail address: yuwenkai@bit.edu.cn}}

\vspace{8pt}
%\noindent{\small A B S T R A C T}

%\noindent\rule{\textwidth}{0.05em}

\noindent{\small \textbf{Abstract:} A secure optical communication requires both high transmission efficiency and high authentication performance, while existing cryptographic key distribution protocols based on ghost imaging have many shortcomings. Here, based on computational ghost imaging, we propose an interactive protocol that enables multi-party cryptographic key distribution over a public network and self-authentication by setting an intermediary that shares partial roles of the server. This fragment-synthesis-based authentication method may facilitate the remote distribution of cryptographic keys.}

%\noindent\rule{\textwidth}{0.05em}

%\noindent {\small\textbf{Keywords}: Optical communications; Secure key distribution; Image reconstruction; Compressive sensing; Single-pixel imaging}

%\vspace{5pt}
%\noindent(Some figures may appear in colour only in the online journal)

%\noindent\rule{\textwidth}{0.05em}

\section{Introduction}
Nowadays, people increasingly rely on the network for daily communications, which makes the data widely spread in the insecure public network and be vulnerable to eavesdropping and tampering by illegal intruders. Therefore, it is particularly important to ensure the security of information transfer process and the accuracy of identity verification. Generally, with inherent capability of parallel processing, the optical encryption systems \cite{Refregier1995,Javidi2000,Situ2005,Meng2006,Zhu2009,Sheng2009,Zhao2015} can efficiently encrypt/decrypt information from multiple dimensions, such as the phase, wavelength, diffraction distance, polarization angle, etc.

As an indirect imaging method, ghost imaging (GI) has attracted a lot of attention in recent years. In the conventional GI scheme \cite{Pittman1995,Gatti2004,Xiong2005,Cheng2009,LiOE2019}, one needs to build a mutually conjugated double-arm optical path: one arm use a pixelated camera to record the time-varying light field of the source; another arm is equipped with a single-pixel (bucket) detector without any spatial resolution to collect the total intensities of the light field that interacts with the object. Then, the profile of the object could be recovered by calculating the intensity correlation function of these two sets of synchronized sampling data. Later, computational ghost imaging (CGI) \cite{Shapiro2008,Bromberg2009} was proposed, it used a spatial light modulator (SLM) to simplify the double-arm light path into one arm, which greatly improved the possibilities of GI's practical applications. During this decade, GI has developed many applications in the fields of lidar \cite{Zhao2012}, microscopy \cite{YuOC2016} and X-ray radiography \cite{Xin2018}. It is worth mentioning that thanks to its random fluctuations of measurements and noisy reconstructed images, GI has become a new favorite in the field of secure optical communication, including image encryption \cite{Clemente2010,Tanha2012}, cryptographic key distribution \cite{SLi2013} and image authentication \cite{Chen2013,YuAO2013,Wu2017,YuAO2019}. However, few GI-based cryptographic key distribution protocols can reduce the sampling ratio while ensuring the efficiency of authentication, without the help of compressed sensing algorithms \cite{Donoho2006,Katz2009,YuOE2014}.

In this work, we propose a protocol for multiparty cryptographic key distribution over a public network and fragment-synthesis-based identity authentication, in a CGI scheme. Compared with the previous works \cite{SLi2013,YuAO2013,YuAO2019}, our approach allows each receiver to simultaneously acquire different cryptographic keys as long as both the server and the receivers are fully trusted via interactive authentication. Additionally, with the help of an intermediary, our scheme can determine whether an attack has occurred.

\section{Principle and Protocol}
As we know, the ghost images can be retrieved by calculating the second-order intensity correlation fluctuation between the speckle reference patterns $I_i$ and the bucket signal $S_{B_i}$:
\begin{align}
\Delta G^{(2)}&=\left\langle(S_{B_i}-\left\langle S_B\right\rangle)(I_i-\left\langle I\right\rangle)\right\rangle\nonumber\\
              &=\left\langle S_BI\right\rangle-\left\langle S_B\right\rangle\left\langle I\right\rangle,
\label{eq:G2}
\end{align}
where $\left\langle u\right\rangle=\frac{1}{N}\sum_{i=1}^Nu^i$ denotes the ensemble average of the signal $u$, $S_{B_i}=\int_{A_l}I_{S_i}(\rho_S)I_i(\rho_A)T(\rho_O)d\rho$, $\rho_S$, $\rho_A$, $\rho_O$ are the spatial coordinates of the source, the speckle reference patterns, and the object, respectively; $I_{S_i}(\rho_S)$ stands for the intensity distribution of the source; $I_i(\rho_A)$ denotes the intensity distribution of speckles; $T(\rho_O)$ refers to the transmission function of the object; $A_l$ presents the integration area; $N$ is the total number of measurements.

\begin{figure}[htbp]
\centering
\includegraphics[width=\linewidth]{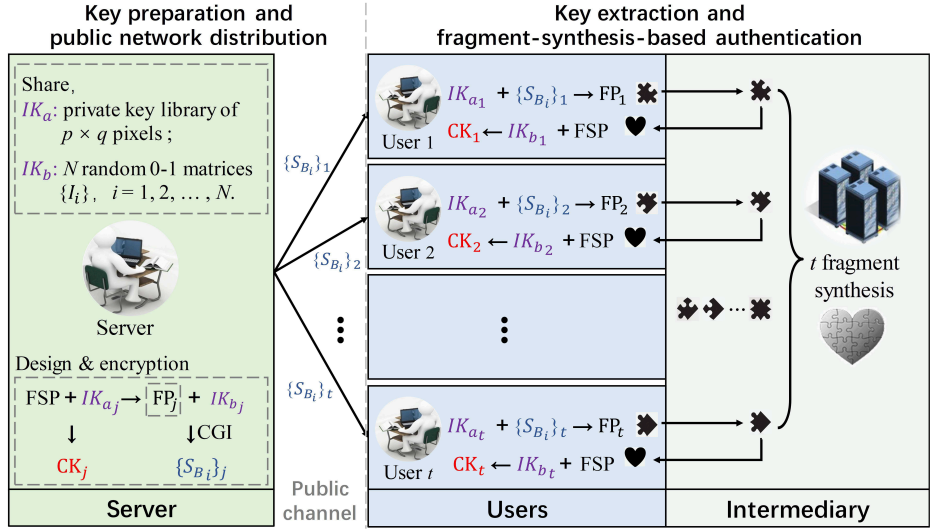}
\caption{Schematic of our fragment-synthesis-based cryptographic key distribution protocol.}
\label{fig:Encrypt}
\end{figure}

The proposed fragment-synthesis-based protocol for $t$ users contains two parts: key preparation and public network distribution, key extraction and fragment-synthesis-based authentication, as shown in Fig.~\ref{fig:Encrypt}.

\textbf{Key preparation and public network distribution}

1). \textbf{Initial keys sharing}. The server needs to share two sets of initial keys (i.e., $IK_a$: a private key library of $p\times q$ pixels and $IK_b$: $N$ random 0-1 matrices $\{I_i\}$, $i=1,2,\cdots,N$) with each receiver through a absolutely secure private channel such as a flash card and a U shield.

2). \textbf{Cryptographic key preparation}. The preparation process is presented in left part of Fig.~\ref{fig:Example}. The server will design a regular binary pattern of $p\times q$ pixels and randomly split it into $t$ binary fragment patterns (FPs) of the same $p\times q$ pixels, which will be used for later fragment-synthesis-based authentication. Referring to the dark pixel positions of regular binary pattern and every private key library (e.g., an unordered alphabet) that has an one-to-one correspondence on each pixel-unit with corresponding FP, the server can easily determine $t$ cryptographic keys (CKs) to be distributed to each legitimate user.

3). \textbf{Encrypted distribution via a public network}. With the help of a CGI setup, the server uses the binary matrices $IK_b$ to randomly sample each binary FP image (as the original object), generating $t$ sequences of single-pixel (bucket) values $\{S_{B_i}\}_j$ ($j=1,2,3,\cdots,t$) in total, which are then sent as encrypted signal carriers through the public network.

\textbf{Key extraction and fragment-synthesis-based authentication}

4). \textbf{Fragment pattern reconstruction}. The users receive the signal $\{S_{B_i}\}_j$ from the public network, and then they are able to reconstruct the respective binary FP with the random binary matrices $IK_b$ shared in advance by calculating the function $\Delta G^{(2)}$.

5). \textbf{Fragment-synthesis-based authentication}. To confirm whether the binary FP recovered by each user is trustworthy requires an authentication. For this purpose, each user should send back his/her FP to the intermediary (as a referee) through a private channel, and the latter can overlay all fragment images together for fragment-synthesis, i.e., joint validation. If the fragment synthesis result is a regular pattern (e.g., geometric pattern), the authentication is successful.

6). \textbf{Key extraction}. After successful authentication, the intermediary will tell the legitimate users the authentication results and return the fragment synthesis pattern (FSP) to the users also through private channels. Then, corresponding to the final FSP, each user can refer to his/her own private key library to obtain the corresponding cryptographic key distributed by the server.

Deserved to be mentioned, since our binary FPs are carefully designed, there will be no overlapped bright pixels in these binary FPs. If there is an authentication error, the intermediary can quickly tell whether the data has been under attack. Note that it is very difficult to accurately locate the insecure user channel, so we discard these distributed keys and replace them with some new ones once the attack occurs.

According to our protocol, we can arrange a server (which transmits the signals through the public network) in a large communication area (such as inter-provincial area, inter-regional area, etc.), and set up some trustworthy intermediaries (which transmits the signals through the private network or local area network) in the small communication areas (such as prefecture-level cities, office buildings, conference rooms, etc.), to achieve multi-party cryptographic key distribution.

\section{Simulation and Experimental Results}
\begin{figure}[htbp]
\centering
\includegraphics[width=\linewidth]{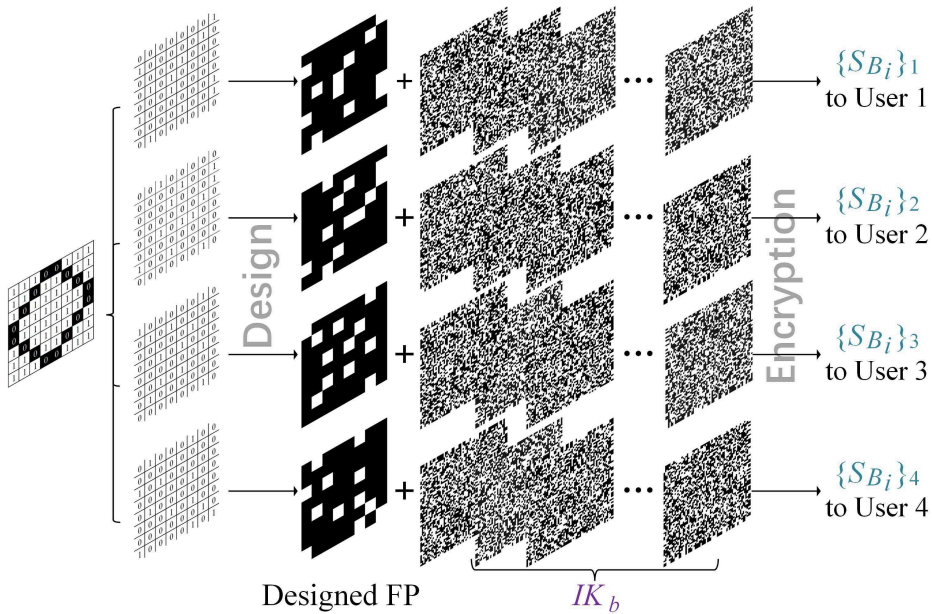}
\caption{Schematic diagram of key preparation and public network distribution procedure, as an instance.}
\label{fig:Example}
\end{figure}

To demonstrate the feasibility of the proposed method, we carry out the numerical simulation. As illustrated in Fig.~\ref{fig:Example}, for simplicity without loss of generality, assume $t=4$, four unordered alphabets of $8\times8$ pixels are generated as the private key libraries $IK_a$ and four sets of 0-1 random matrices are used as $IK_b$. Then, the server will elaborately design a regular binary pattern of pixel-size $8\times8$, such as a rhombus as shown in Fig.~\ref{fig:Example}. Next, the server can split the regular binary image into four irregular $0/1$ FPs of the same $8\times8$ pixels. Since the images recovered by second-order intensity correlation function has unavoidable noise fluctuations, we apply an upsampling strategy, i.e., set $\nu\times\nu$ pixels of the 0-1 random matrix correspond to one pixel-unit of the FP or that of the unordered alphabet. This is equivalent to enlarging the fragment image $\nu$ times horizontally and vertically. In this simulation, we set $\nu=8$, so the pixel size of each random matrix is $64\times64$, as shown in Fig.~\ref{fig:Example}.

\begin{figure}[htbp]
\centering
\includegraphics[width=\linewidth]{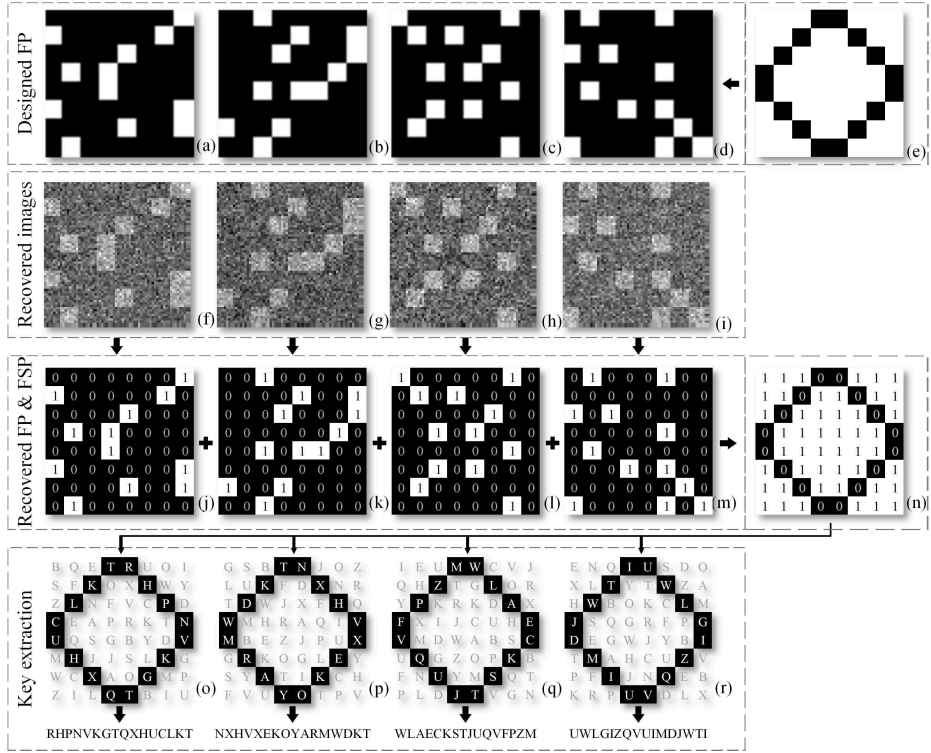}
\caption{Simulation results. (a)--(d) are the four designed fragment images; (e) is the original FSP; (f)--(i) are the results recovered by GI, and the key extraction method is illustrated in (i); (j)--(m) are four binarized matrices obtained from (f)--(i); (n) is the FSP calculated from (j)--(m); (o)--(r) show the extraction process of distributed cryptographic keys according to four unordered alphabets.}
\label{fig:SimResults}
\end{figure}

The simulation results with the number of measurements $N=4096$ are presented in Fig.~\ref{fig:SimResults}. The designed fragment images are given in Figs.~\ref{fig:SimResults}(a)--\ref{fig:SimResults}(d), and the original fragment synthesis image is shown in Fig.~\ref{fig:SimResults}(e). In the key extraction process of the previous works \cite{SLi2013,YuAO2013}, some digits after the decimal point of each gray value of the recovered ghost image are extracted to form a bit sequence as the distributed cryptographic key, thus a lot of measurements are needed to ensure the randomness of the distributed key. Different from this method, here we smooth the recovered image (Figs.~\ref{fig:SimResults}(f)--\ref{fig:SimResults}(i)) to get corresponding binary images, as shown in Figs.~\ref{fig:SimResults}(j)--\ref{fig:SimResults}(m). By superimposing the pixel values of these four binarized matrices (i.e., retrieved binary FPs), we can acquire the FSP (see Fig.~\ref{fig:SimResults}(n)). Here, the result is perfect, showing a regular geometric pattern, so we can confirm that there is no attack and all the users are legal. The users refer to their unordered alphabets to extract the distributed cryptographic keys, as illustrated in Figs.~\ref{fig:SimResults}(o)--\ref{fig:SimResults}(r). Since the binary FPs are carefully designed with no overlap between each other, the FSP should also be a matrix consisting of 0 and 1. If the recovered FSP has any pixel-unit with a value greater than 1, then the intermediary can tell whether the data transmission over the public network is under attack.

\begin{figure}[htbp]
\centering
\includegraphics[width=\linewidth]{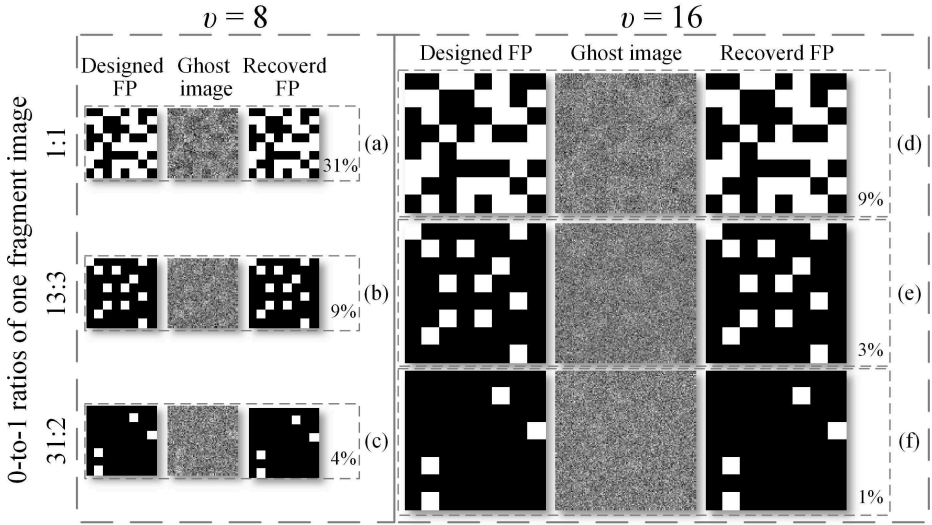}
\caption{The subsampling limits for the binarization of recognizable recovered fragment images. (a)--(c) and (d)--(f) are three original fragment images with 1:1, 13:3 and 31:2 0-to-1 ratios, corresponding ghost images of $\Delta G^{(2)}$, and the binarized images via sorting, for $\nu=8$ and 16, respectively.}
\label{fig:Samping}
\end{figure}

When the sampling rate is very low, the binarization process can also be realized by using a sorting method. To be specific, we set every $\nu\times\nu$ pixels of the recovered images (the random matrix for the same) as a pixel-unit, and average all the pixel values in it, then sort all mean values of one reconstructed image in a descending order. By this means, if the legitimate user knows the total number $M_j$ of bright pixels in his/her fragment image in advance, he/her can easily light up the pixel units where the first $M_j$ means are located, on a matrix of all zeros. To verify this sorting method, we perform some other simulations as given in Fig.~\ref{fig:Samping}. Here, we set three 0-to-1 ratios of one fragment image, they are 1:1, 13:3 and 31:2, respectively. For an amplification coefficient of one pixel $\nu$, the actual resolution of the recovered images become $p\nu\times q\nu$. Due to the unavoidable noise fluctuations existing in ghost images, directly use $p\times q$ modulated matrices for GI reconstruction, the binarization process (smoothing or sorting) will fail to work. But if we use $p\nu\times q\nu$ modulated matrices for GI reconstruction, the number of measurements for recovering distinguishable ghost images will be more than ten times the actual $\eta=p\nu\times q\nu$ pixels. According to our previous work \cite{YuOE2014}, the probability of the reconstructed pixel values locating in the pixel region of the same original gray value obeys a Gaussian distribution. Here, each pixel of the original fragment image will be equivalently scaled to a $\nu\time\nu$ pixel-unit. By using the mean of recovered pixel values in these pixel-units can help reduce the total number of measurements. In our simulations, when $\nu=8$, the subsampling ratio limits for binarizing recognizable reconstructed fragment images are about $31\%$, $9\%$ and $4\%$ of $64\times64=4096$ (see Figs.~\ref{fig:Samping}(a)--\ref{fig:Samping}(c)); when $\nu=16$, the latters are about 9\%, 3\% and 1\% of $128\times128=16384$ (see Figs.~\ref{fig:Samping}(d)--\ref{fig:Samping}(f), where some details are even hard to be resolved). It is clear that as the 0-to-1 ratio increases, the sampling ratio will be reduced. All above sampling ratio limits were obtained from 100 repeated experiments.

\begin{figure}[htbp]
\centering
\includegraphics[width=\linewidth]{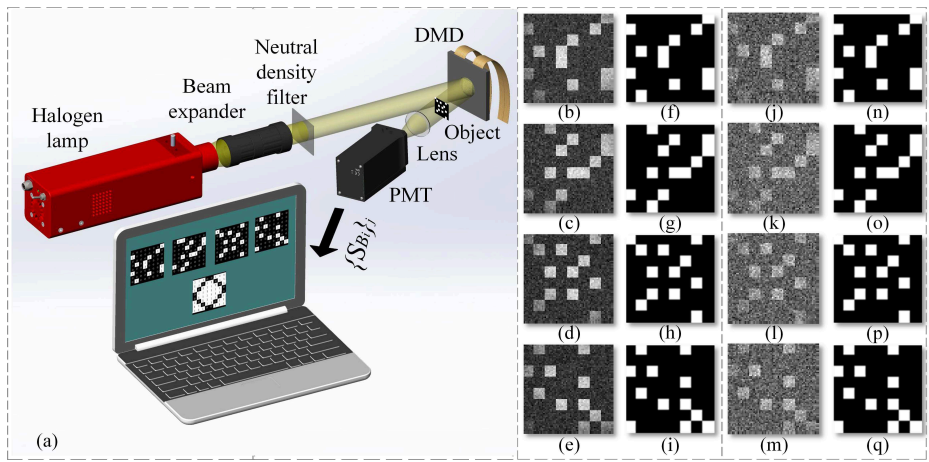}
\caption{Experimental setup (a) and results (b-i) with $\nu=8$. (b)--(e) and (j)--(m) show the four recovered results with $N=20480$ and 4096 measurements, respectively. (f)--(i) and (n)--(q) give the binarized results of (b)--(e) and (j)--(m) by using the smoothing and sorting methods, respectively.}
\label{fig:Experiment}
\end{figure}

In experiment, we used a CGI setup to demonstrate this scheme, as shown in Fig.~\ref{fig:Experiment}. The light beam from the halogen lamp is amplified, collimated, and attenuated to form a parallel beam which then illumines the working plane of the digital micromirror device (DMD). Then, the reflected light from the DMD passes through a convergent lens (of focal length 50~mm) and is collected into a photomultiplier tube (PMT). We set $\nu=8$, and encode the 0-1 matrices $IK_b$ of $64\times64$ are encoded onto the DMD for random sampling. The frame rate of the DMD is set to 500~Hz. For four designed fragment images (as the original images), there will be four sequences of single-pixel values $\{S_{B_i}\}_j$ ($j=1,2,3,4$). The GI reconstructed results are presented as Figs.~\ref{fig:Experiment}(b)--\ref{fig:Experiment}(e) and \ref{fig:Experiment}(j)--\ref{fig:Experiment}(m), and their binarized results are shown in Figs.~\ref{fig:Experiment}(f)--\ref{fig:Experiment}(i) and \ref{fig:Experiment}(n)--\ref{fig:Experiment}(q), with the number of measurements $N=20480$ and 4096, by using the smoothing and sorting methods, respectively. Although the image qualities of GI with 4096 measurements are far worse than those with 20480 measurements (oversampling), we can still obtain correct FP results from such a small amount of measurements by using the sorting method. For this reason, we repeat the analysis did in simulation and acquire the subsampling ratio limit of this experimental data applying the sorting method, about 7\% (286 measurements). This limit is just a little lower than the simulated value 9\%, due to the unevenness of light field, which causes the central bright pixel-units to be more prominent (conducive to sorting extraction).

\section{Discussion}
\begin{figure}[htbp]
\centering
\includegraphics[width=\linewidth]{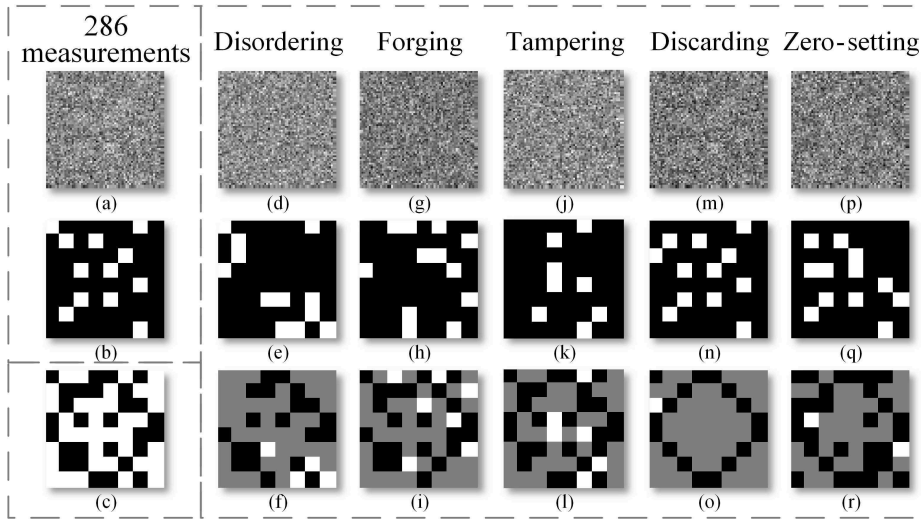}
\caption{Results in presence of different attacks. (a)--(b) are the recovered image and the resolved FP as references using unattacked $\{S_{B_i}\}_3$ of length 286 which is distributed to User 3; (c) is a fragment-synthesis intermediate result using only three correct FPs of Users 1, 2, 4; (d)--(f), (g)--(i), (j)--(l), (m)--(o) and (p)--(r) are the recovered ghost images, corresponding binarized images and the final fragment-synthesis results (adding these wrong binarized image of User 3 with the computed correct intermediate result (c)) under five types of attacks, i.e. disordering, forging, tampering, discarding and zero-setting.}
\label{fig:Tamper}
\end{figure}

Next, we will analyze the security of this protocol under various attacks. Since only the sequences of single-pixel values are transmitted in the public network, the signal $\{S_{B_i}\}_j$ is the only place that can be attacked by the eavesdropper Eve. Without loss of generality, we take the sequence $\{S_{B_i}\}_3$ experimentally measured for User 3 as the target of attack, and assume that Eve cannot get any initial keys. Here, the used subsampling ratio for binarizing recognizable recovered FP is 7\% (the obtained limit), so the length of $\{S_{B_i}\}_3$ is 286. The reconstructed ghost image and FP using this unattacked $\{S_{B_i}\}_3$ are given in Figs.~\ref{fig:Tamper}(a) and \ref{fig:Tamper}(b) as references. In Fig.~\ref{fig:Tamper}(c), we also provide the fragment-synthesis intermediate result that only uses three correct FPs of Users 1, 2, 4. Here, we test the authentication results in presence of the following five types of attacks: disordering the sequence, forging a new sequence to replace the original one, changing some values (1\% of 286) in the sequence, discarding some data in the sequence (1\% of 286) and setting some part (also 1\% of 286) of the data to zero. Their recovered ghost images, corresponding binarized images and the final fragment-synthesis results (combined with the correct intermediate result Fig.~\ref{fig:Tamper}(c)) are presented in Figs.~\ref{fig:Tamper}(d)--\ref{fig:Tamper}(f), \ref{fig:Tamper}(g)--\ref{fig:Tamper}(i), \ref{fig:Tamper}(j)--\ref{fig:Tamper}(l), \ref{fig:Tamper}(m)--\ref{fig:Tamper}(o) and \ref{fig:Tamper}(p)--\ref{fig:Tamper}(r), respectively. As we can see, compared with Fig.~\ref{fig:Tamper}(b), there present various errors in Figs.~\ref{fig:Tamper}(e), \ref{fig:Tamper}(h), \ref{fig:Tamper}(k), \ref{fig:Tamper}(n) and \ref{fig:Tamper}(q), and the final superimposed results as shown in Figs.~\ref{fig:Tamper}(f), \ref{fig:Tamper}(i), \ref{fig:Tamper}(l), \ref{fig:Tamper}(o) and \ref{fig:Tamper}(r) are no longer binary images, telling that the transmitted data has been attacked.

\section{Conclusion}
In conclusion, a fragment-synthesis-based protocol for multiparty cryptographic key distribution over a public network with an authentication capability is proposed. By setting an intermediary, this protocol can quickly judge whether the transmitted data has been under attack, via FSP. Both numerical simulation and experimental results have demonstrated the feasibility of this protocol, and the performance under different attacks has also been discussed. This protocol may offer a new way for applying the intermediary in optical secure communication, and will benefit many practical security applications.

\section*{Acknowledgements}
This work is supported by the National Natural Science Foundation of China (61801022); the National Key Research and Development Program of China (2016YFE0131500); and the Civil Space Project of China (D040301).

\end{document}